# Amplitude-Phase Fusion for Enhanced Electrocardiogram Morphological Analysis

**Shuaicong Hu**[1], **Yanan Wang**[1], **Jian Liu**[1], **Jingyu Lin**[2], **Shengmei Qin**[2,3], **Zhenning Nie**[2,3], **Zhifeng Yao**[2,3], **Wenjie Cai**[4], and **Cuiwei Yang**[1,5]

[1]Department of Biomedical Engineering, School of Information Science and Technology, Fudan University, Shanghai, 200433, China
[2]Shanghai Institute of Cardiovascular Diseases, Zhongshan Hospital, Fudan University, Shanghai 200032, China
[3]Department of Cardiology, Zhongshan Hospital, Fudan University, Shanghai 200032, China
[4]School of Health Science and Engineering, University of Shanghai for Science and Technology, Shanghai 200093, China
[5]Key Laboratory of Medical Imaging Computing and Computer Assisted Intervention of Shanghai, Shanghai 200093, China
schu22@m.fudan.edu.cn and yangcw@fudan.edu.cn

*Abstract*—Considering the variability of amplitude and phase patterns in electrocardiogram (ECG) signals due to cardiac activity and individual differences, existing entropy-based studies have not fully utilized these two patterns and lack integration. To address this gap, this paper proposes a novel fusion entropy metric, morphological ECG entropy (MEE) for the first time, specifically designed for ECG morphology, to comprehensively describe the fusion of amplitude and phase patterns. MEE is computed based on beat-level samples, enabling detailed analysis of each cardiac cycle. Experimental results demonstrate that MEE achieves rapid, accurate, and label-free localization of abnormal ECG arrhythmia regions. Furthermore, MEE provides a method for assessing sample diversity, facilitating compression of imbalanced training sets (via representative sample selection), and outperforms random pruning. Additionally, MEE exhibits the ability to describe areas of poor quality. By discussing, it proves the robustness of MEE value calculation to noise interference and its low computational complexity. Finally, we integrate this method into a clinical interactive interface to provide a more convenient and intuitive user experience. These findings indicate that MEE serves as a valuable clinical descriptor for ECG characterization. The implementation code can be referenced at the following link: https://github.com/fdu-harry/ECG-MEE-metric.

## 1. INTRODUCTION

ENTROPY is a physical quantity introduced with the Second Law of Thermodynamics [1], used to measure the degree of disorder or uncertainty in a system. In the early 1940s, Shannon introduced the concept of entropy as a measure of information in information theory [2], which quantifies the amount of useful knowledge contained in information.

The electrocardiogram (ECG) is a widely utilized diagnostic tool for cardiovascular disease. Numerous methods have been developed to analyze ECG using entropy analysis, including the application of entropy-based metrics to study atrial fibrillation, other arrhythmias, and disparities between heart disease patients and healthy populations [3, 4]. Additionally, some researchers employ permutation entropy (PE) to identify abnormal heart rates [5, 6]. However, current entropy measurement techniques for ECG often fail to fully incorporate both amplitude and phase information, which are crucial for accurately assessing the morphology of ECG signals [7-10]. There is a gap in current research, necessitating the development of a novel entropy measurement method capable of more effectively assessing the patterns and latent information present in ECG signals. This work explores an entropy-based metric specifically designed for ECG, considering two factors: amplitude and phase patterns. The aim is to uncover hidden patterns within the distribution of signal amplitudes and the temporal dimension of ECG local waveforms, which encapsulate physiological information.

By simultaneously considering both the amplitude and phase information, the proposed metric enhances the capture of patterns and hidden information in ECG morphology. This innovative approach has the potential to enable low-cost, unsupervised, and rapid screening for arrhythmias, thus improving diagnostic efficiency. Furthermore, the method also holds promise for measuring signal diversity. By compressing redundant samples in the dataset and enhancing representation, it facilitates the construction of diverse datasets. Building upon this foundation, it will aid in the development of more robust and interpretable machine learning (ML) models for ECG analysis. Leveraging the superior characterization capabilities of the proposed entropy-based measure for ECG morphology, we designate it as Morphological ECG Entropy (MEE). The distinctiveness of MEE lies in its capacity to screen for arrhythmias and evaluate sample diversity at the beat level, while exhibiting high sensitivity to noisy regions. Its main contributions are as follows:

- This paper designs two entropy measurement methods for ECG amplitude and phase patterns. Through their complementary integration, the MEE is formed, and various types of experiments validate its outstanding potential as a morphological assessment measure at the beat level.
- The methodologies for applying the MEE metric at the beat level of single-lead and standard 12-lead ECG signals are provided, showcasing their value in: 1) Unsupervised rapid screening for morphological abnormal arrhythmias, 2) Constructing representative sample sets to facilitate richer representation learning and performance enhancement, 3) Sensitivity to noise regions. Moreover, it offers inspiration and opportunities for further

- advancements to future researchers.
- Experimental evidence demonstrates the robustness under noise interference and high computational efficiency of MEE. Furthermore, we integrate the MEE metric into a clinical tool for physician use, showcasing its practical value and potential to become a widely adopted metric standard in ECG morphological assessments.

The rest of this paper is organized into five sections. Section 2 introduces the motivation of the proposed MEE metric. Section 3 describes the methodology. Section 4 presents the experimental results. Section 5 presents the discussion part. Section 6 concludes this work.

## 2. MOTIVATION

Research on one-dimensional ECG signals is extensive. Voltage variations are formed by recording the sequence of potential values at a given sampling rate, which serves as the cornerstone for detecting arrhythmias and other diseases. Heartbeats, as the fundamental units of ECG signals, have their R-peak positions localized by well-established, high-precision detection algorithms, which function as periodic reference coordinates. However, existing research lacks fully exploited measurement methods derived from the morphological patterns of this ECG basic component. Although numerous entropy-based metrics have been proposed, they primarily focus on quantifying regularities in short time series based on local similarity or ordinal patterns. For instance, Approximate Entropy (AE) [11] divides the signal into different lengths of subsequences, computes the similarity between these subsequences, and then calculates the AE based on the similarity values. However, AE may produce biased estimates for self-matching signals. To address this issue, Sample Entropy (SE) [10] is introduced to avoid self-comparisons during computation and mitigate the meaningless logarithmic term. Studies have shown that both AE and SE exhibit poor statistical stability. To further address the shortcomings of hard partitioning induced by the two-state classifier characteristics of the Heaviside function, Fuzzy Entropy (FE) [12] introduces the concept of membership to better adapt to the fact that boundaries between classes in the real world are often fuzzy. PE [5] computes measures by statistically analyzing the occurrence of permutation patterns in a time series, which has been utilized in heartbeat detection. However, it is foreseeable that permutation patterns will be confused in the presence of noise interference, leading to instability in the computed results.

Although the classical entropy methods mentioned above have been continuously improved, they still do not fully consider the characteristics of ECG and cannot be used as a dedicated metric for ECG signals. To address this, Liu et al. [4] proposed a new entropy measure, namely Fuzzy Measure Entropy (FuzzyMEn), for clinical heart rate variability (HRV) analysis. By correcting abnormal intervals in locating R-peak values through the detection algorithm and deriving entropy values based on R-R intervals (the time between successive R peaks), FuzzyMEn explores rhythm pattern information and demonstrates the ability to distinguish between healthy individuals and heart failure patients. However, the morphological features of heartbeats, which serve as the fundamental units of ECG, have been overlooked by researchers, and there is still a lack of comprehensive exploration of beat-level morphology information. Morphological information is crucial in clinical diagnosis, where differences in waveform patterns such as P waves, QRS complexes, and ST segments reflect various cardiac activities, while the timings of different wavelet occurrences provide information on electrical conduction differences. After fully considering the aforementioned key issues, this paper proposes two metrics, Bandwidth Entropy (BWE) and Wavelet set Entropy (WSE), to describe amplitude and phase patterns, respectively. By complementarily integrating these patterns, MEE emerges for morphological assessment at the beat level. This index directly adapts to the morphological structure of the ECG signal, presenting broad application prospects, primarily encompassing the following conceptual applications:

1) Clinical settings generate vast amounts of ECG data on a daily basis, yet doctors are unable to annotate all data, and fine-grained arrhythmia exclusion relies solely on expert visual inspection of heartbeats. Current ML models, including well-known deep learning (DL) models, rely on extensive annotated data for training and lack robustness on independent test sets. Moreover, these models suffer from poor interpretability, leading to inherent mistrust among clinical practitioners, ultimately requiring doctor calibration of diagnostic results. The MEE, proposed in this paper specifically for ECG signals, aims to transform multidimensional feature values of multiple sampling points into a single computable real value at the beat level. By compressing ECG sequences into a single value, it provides expert-consistent heartbeat trend results across multiple heartbeat perspectives (fluctuations in MEE values for abnormal heartbeats, while sinus heartbeats remain stable). The calculation of this index requires no labeling and is solely based on the signal's time series. Its goal is to provide a rapid, accurate, and label-free method for arrhythmia screening, significantly reducing the workload of doctors.

2) Dataset pruning and representative sample selection aid in improving model training speed. In existing works, representative sample selection is commonly employed to train models, reducing redundant features from majority class samples, which helps mitigate the impact of class imbalance issues on model learning useful minority class features. Additionally, it helps reduce training time and enhances model generalizability by training from scratch.

3) In existing ML algorithms, data grouping is random and fails to fully leverage the data-centric approach, maximizing the utilization of existing annotated data. The calculation of the MEE index helps identify, before the model training process, the group with the highest variance of MEE index within similar samples, thereby enhancing the model's generalization ability and avoiding the randomness of grouping.

4) Calculating MEE values aids in capturing noise signal segments to eliminate noise features while retaining class-related features.

## 3. METHODOLOGY

This paper comprehensively evaluates four classical entropy measurement methods: PE, AE, SE, and FE. Additionally, it introduces three new entropy measures proposed in this study: BWE, WSE, and MEE. Brief descriptions of existing metrics

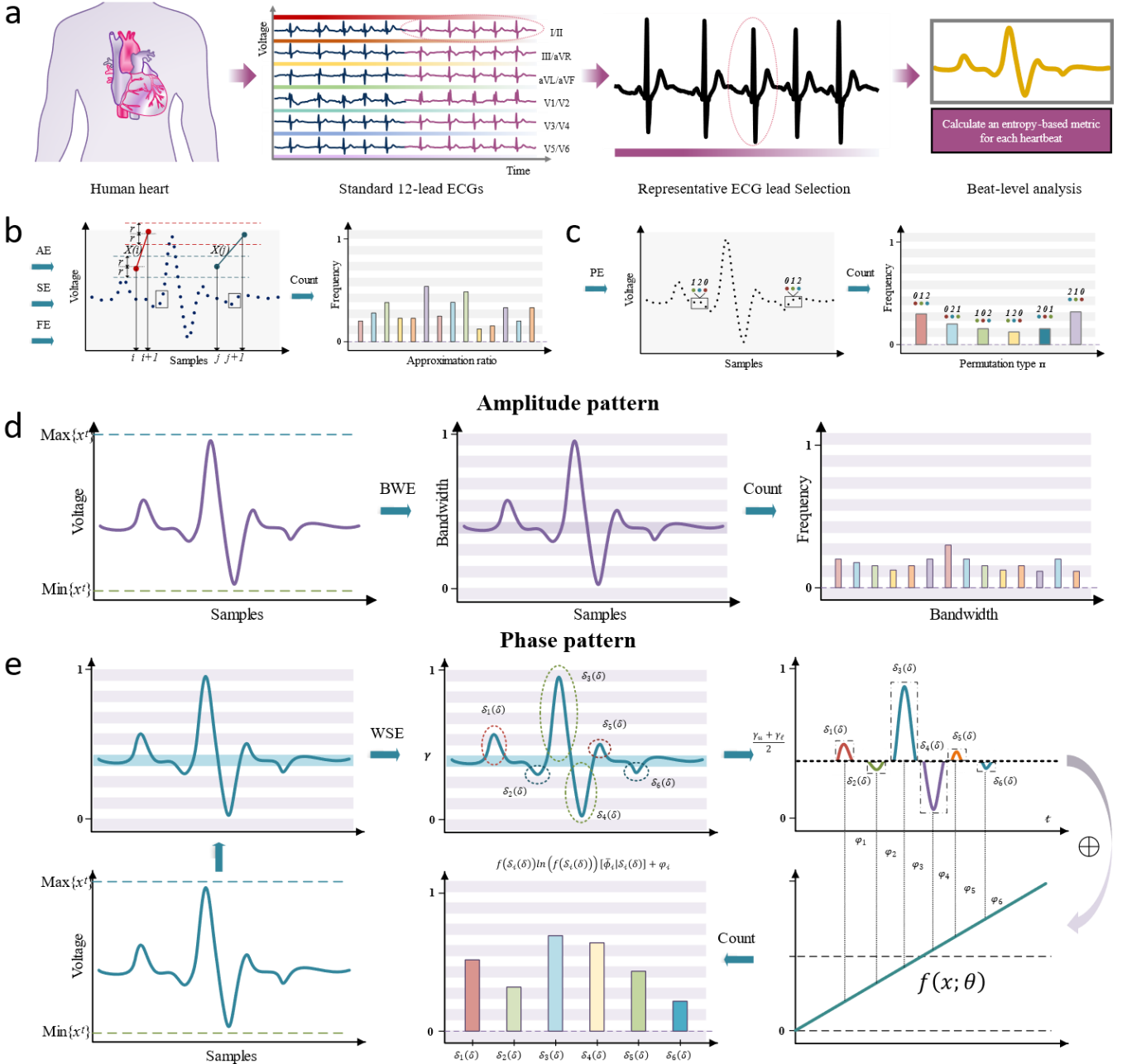

**Fig. 1.** Different levels of ECG analysis and various entropy-based measures are employed. (a) The cardiac electrical activity signal is acquired sequentially in standard 12-lead configuration, representative lead II, and individual heartbeats. (b) The implementation details of AE, SE, and FE primarily rely on statistical analyses of similar patterns, progressively serving as a set of evolved methodologies. (c) The implementation details of PE primarily aim to capture sequential patterns within the sequences. (d) Bandwidth Entropy (BWE) partitions the amplitude range and computes the probability of amplitude samples occurring within different ranges. Additionally, bandwidth-associated amplitudes are assigned coefficients for each entropy term to enhance specificity across different regions. (e) Wavelet Set Entropy (WSE) localizes approximate baselines and extracts consecutive sets of characteristic waves, which lie either above or below the baseline. Energy is allocated for each wavelet set, mapped to phase-generated bias terms, culminating in entropy metrics.

and detailed explanations of the proposed metrics are provided below.

### 3.1. Permutation Entropy (PE)

For a given time series $\{x_t |_{t=0,1,...,T-1}\}$ of length $T$, the PE [13, 14] is the Shannon entropy of the permutation type $\pi$, and order $n$ denotes the number of sample points to be arranged. The expression of the order $n$ PE is:

$$H(n) = -\sum_n p(\pi) \ln(p(\pi)) \quad (1)$$

where $p(\pi)$ denotes relative frequency of permutation type $\pi$, and can be expressed by:

$$p(\pi) = \frac{\#\{t \mid t \leq T-n, (x_{t+1},...,x_{t+n}) \text{ has type } \pi\}}{T-n+1} \quad (2)$$

It can be seen from the above formula that PE can be used as an information measure of the frequency of occurrence of various permutations in a time series, which may be effective for measuring the abundance of sorting information contained in an ECG sequence.

### 3.2. Approximate Entropy (AE), Sample Entropy (SE), and Fuzzy Entropy (FE)

Due to the convergence of AE, SE, and FE as a continuum of improvement methods, we will present them collectively. For AE, its computational expression is as follows:

$$H_{AE}(m,r) = \phi^m(r) - \phi^{m+1}(r) \tag{3}$$

where $\phi^m(r) = \frac{1}{N-m+1}\sum_{i=1}^{N-m+1}\ln C_i^m(r)$, $C_i^m(r)$ represents the ratio of the approximate count within a tolerance $r$ to the total count for the $i$ th local vector with a pattern dimension of $m$. By iterating through the signal, $\phi^m(r)$ is obtained, and by incrementing the pattern dimension $m$ by 1, this process is repeated, resulting in the subtraction to derive the AE. Since the approximation of local vectors to themselves in AE introduces computational bias, SE is proposed as an improvement method:

$$H_{SE}(m,r) = \ln B^m(r) - \ln B^{m+1}(r) \tag{4}$$

where $B^m(r) = \frac{1}{N-m}\sum_{i=1}^{N-m} B_i^m(r)$, in this method, the $i$ th local vector does not require approximation to itself, and logarithmic computation is not applied to the ratio of approximate count to total count. This avoids situations where the ratio being zero leads to meaningless outcomes. Instead, logarithmic computation is applied to $B^m(r)$. Furthermore, since both AE and SE are based on a similarity definition that relies on a binary classifier, this hard partitioning pattern does not fit well with the fuzzy class boundaries present in the real world. Therefore, FE improves the computation by introducing membership degrees:

$$H_{FE}(m,n,r) = \ln A^m(n,r) - \ln A^{m+1}(n,r) \tag{5}$$

where $A^m(n,r) = \frac{1}{N-m}\sum_{i=1}^{N-m}\left(\frac{1}{N-m-1}\sum_{j=1,j\neq i}^{N-m} D_{ij}^m\right)$, in the computation process, the concept of membership degree is introduced, which is calculated based on the maximum difference between two local vector endpoints derived from the signal, denoted as $d_{ij}^m$, to avoid the hard partitioning issue caused by the tolerance $r$. The membership degree is denoted as $D_{ij}^m = \exp\left(-\left(d_{ij}^m\right)^n / r\right)$, where parameter $n$, serving as the weight for fuzzy vector similarity, is typically set to 2 or 3. Unlike AE and SE, local vectors need to subtract their mean value.

Since the aforementioned three measurement methods are widely used in the field of biological signal detection, they serve as classical baseline methods for evaluating the metrics proposed in this paper.

### 3.3. Proposed Bandwidth Entropy (BWE)

Considering that the potential information contained in the ECG signal is usually reflected in the shape and amplitude of the P-wave, T-wave, and QRS-complex, we first count the ECG sampling points in different amplitude ranges, and the relative frequency of the sampling points in the corresponding amplitude bandwidth under the statistical sense is regarded as the probability, and the Shannon entropy is calculated. To ensure the specificity of each bandwidth, we multiply the contribution of each bandwidth to the Shannon entropy by the median value of the bandwidth. This approach helps alleviate situations where signals with significant morphological differences may obtain similar entropy values.

Considering the difference between the amplitudes of ECG signals, to make all the signals can be counted in the same super-bandwidth, we normalize the ECG sequence $\{x_t\}_{t=0,1,\ldots,T-1}$ of length $T$ [15]:

$$y_t = \frac{x_t - \min\{x_t\}}{\max\{x_t\} - \min\{x_t\}} \tag{6}$$

And the proposed BWE can be calculated by:

$$H_{BWE} = -\sum_i M(B_i(\delta)) f(B_i(\delta)) \ln(f(B_i(\delta))) \tag{7}$$

where $B_i(\delta)$ is the sampling point set within the $i$ th bandwidth we divided, $f(B_i(\delta))$ is the relative frequency of the $i$ th bandwidth point set on the whole section of the signal:

$$f(B_i(\delta)) = \frac{B_i(\delta)}{\sum_i B_i(\delta)} \tag{8}$$

where $M(\cdot)$ is the median of the bandwidth range, $H_{BWE}$ can be seen as Shannon entropy with bandwidth location information. In our study, we defined this entropy measure as a candidate method for evaluating the data diversity of ECG signals.

### 3.4. Proposed Wavelet Set Entropy (WSE)

We consider the following objective facts: the cardiac electrophysiological change process corresponding to the ECG signal is the most fundamental reason for generating information, which is reflected as a continuous set of wavelets on the ECG, and the fluctuations in the range near the baseline can be regarded as no information participation, therefore, we propose another entropy measure for ECG signals.

We first normalized the ECG signal so that the wavelet set has the same unit of measure, then divided the amplitude domain in equal proportions, and regarded the bandwidth with the most sample points as the approximate baseline position. Then, the wavelet sets on both sides of the upper and lower bounds of the baseline bandwidth are counted, and the points on the same side with no less than 3 consecutive points are regarded as a wavelet set, and the relative frequencies of the wavelet sets are counted to calculate Shannon entropy. Also similar to ECG-bandwidth entropy, the amplitude mean of the wavelet set is multiplied before the Shannon entropy sub-term to carry specificity information. Considering the sequential appearance of wavelet sets in the time domain, which encapsulates the electrical conduction time information of the heart, we thus append bias $\varphi_i$ to each wavelet set $S_i(\delta)$. The resulting WSE is obtained:

$$H_{WSE} = -\sum_i \frac{S_i(\delta)}{\sum_i S_i(\delta)} \ln\left(\frac{S_i(\delta)}{\sum_i S_i(\delta)}\right) \left[\bar{\phi}_i | S_i(\delta)\right] + \varphi_i \quad (9)$$

$$S_i(\delta) \text{ satisfy } num(\delta) \geq 3 \quad (10)$$

where $S_i(\delta)$ is the sampling point set of $i$ th wavelet set, $\left[\bar{\phi}_i | S_i(\delta)\right]$ denotes the average amplitude of the whole sampling points in the $i$ th wavelet set $S_i(\delta)$:

$$\left[\bar{\phi}_i | S_i(\delta)\right] = \begin{cases} \dfrac{\sum_\delta S_i(\delta) - \dfrac{\gamma_u + \gamma_l}{2}}{\sum \delta}, & \text{iff } S_i(\delta) \geq \gamma_u \\ \dfrac{-\sum_\delta S_i(\delta) + \dfrac{\gamma_u + \gamma_l}{2}}{\sum \delta}, & \text{iff } S_i(\delta) \leq \gamma_l \end{cases} \quad (11)$$

where $\gamma_u$, $\gamma_l$ denote the upper bound and lower bound of the baseline bandwidth, respectively. Moreover, $\varphi_i$ is a bias for the $i$ th wavelet set $S_i(\delta)$ which used to apply phase information:

$$\varphi_i = f\left(\frac{\left[\bar{\phi}_i | S_i(\delta)\right] \langle P_i | S_i(\delta) \rangle}{T}; \theta\right) \quad (12)$$

$$P_i = idx\left(where \sum_o^{P_i} S_i(\delta) = \frac{1}{2} A(S_i(\delta))\right) \quad (13)$$

In this study, $P_i$ represents the x-coordinate corresponding to the area bisector of the $i$ th wavelet set $S_i(\delta)$. It approximates the time when the total energy of $S_i(\delta)$ appears on the time axis. This time is normalized based on the length of the heartbeat hierarchy sequence to assess relative timing. Additionally, energy weights are assigned to moments through coefficient $\left[\bar{\phi}_i | S_i(\delta)\right]$. Here, $f(x;\theta)$ represents a mapping function used to map moments to other value ranges. In this paper, we provide a linear form $f(x;\theta) = 10x$ based on empirical evidence, ensuring the accumulation of temporal information without cancellation.

Additionally, we notice that due to the energy weighting factor $\left[\bar{\phi}_i | S_i(\delta)\right]$ applied to each wavelet set $S_i(\delta)$, negative sub-terms will be obtained below the baseline and cancel out the positive sub-terms from the wavelet sets above the baseline. Therefore, we consider two scenarios, namely distinguishing positive and negative sub-terms, and converting negative sub-terms to positive ones (without distinguishing the direction of wavelet sets above and below the approximate baseline), naming them WSE-I and WSE-II, respectively.

## 3.5. Fusing amplitude (BWE) and phase information (WSE) - Morphological ECG entropy (MEE)

In order to combine the entropy index with both amplitude and phase information of a signal, we also aim to represent the index as a real number instead of a complex number. Since complex indices would make it difficult to screen for abnormal signals, as we cannot determine the range of complex indices for normal signal segments, and therefore cannot provide reference indices for abnormal signal screening. Therefore, we are inspired by Euler's formula to improve it in the real number domain for assigning an index to ECG signals that integrates both amplitude and phase information [16]:

$$\begin{aligned} e^{i\pi} + 1 &= 0, \\ e^{i\theta} &= \cos\theta + i\sin\theta \end{aligned} \quad (14)$$

Euler's formula is an important mathematical formula that relates three fundamental constants: $e$, $\pi$, and the imaginary unit $i$. It has wide applications in the field of signal processing because it can represent complex numbers in the form of exponential functions, which facilitates signal processing. In the field of signal processing, it is important to consider both the phase and amplitude information of signals. For example, Euler's formula can be utilized to calculate the Fourier transform of signals, which is crucial in fields such as signal processing and communication [17]. In some applications, however, we may not need to utilize complex metrics, as complex calculations can increase computational complexity. To avoid this situation, we can draw inspiration from Euler's formula and its derived version to design a metric that combines amplitude and phase information in the real number domain for computation:

$$H_{MEE} = H_{BWE} \times e^{\frac{H_{WSE}}{2\pi}} \quad (15)$$

In this formula, $H_{BWE}$ and $H_{WSE}$ represent amplitude pattern and phase pattern, respectively. By combining them into a single real number, we obtain the final MEE $H_{MEE}$, which provides a reference measure for subsequent screening of arrhythmias and diversity assessment. Moreover, compared to the complex form, it has a lower computational cost. We also consider another convenient fusion method, computing the sum of squares of both and then averaging them:

$$H_{MEE} = \frac{H_{BWE}^2 + H_{WSE}^2}{2} \quad (16)$$

The squaring operation amplifies the differences between BWE and WSE indices to enhance the detectability of outliers, while the averaging operation allows both indices to simultaneously participate in morphological measurements.

## 3.6. Proposed Bandwidth Picking Algorithm

Since the MEE-based approach is label-free, an automatic method needs to be designed to process the calculated entropy metric for purposes such as arrhythmia screening and diversity assessment. In this paper, we propose an entropy metric processing method, which termed the bandwidth picking method. For the arrhythmia screening task, the sinus beats have a stable morphology and their corresponding MEE metric has a close and stable value. Based on the above knowledge, we set the maximum and minimum values of the MEE metric as the upper and lower bounds, respectively, and divide all values into several equally spaced bandwidths, set to 40 in this paper. Then we determine the bandwidth with the largest number of samples, and the MEE metric value corresponding to this bandwidth can be regarded as the reference value. By comparing the MEE value sequence with the reference value, heartbeats with a fluctuation ratio greater than a threshold $\alpha$ are considered abnormal. The fluctuation ratio is defined as follows:

$$f(k) = \frac{\left| M(k) - M_{ref}(k) \right|}{M_{ref}(k) + \sigma_M} \tag{17}$$

$$M_{ref}(k) = Mid \left[ \arg\max\nolimits_{num}(band_{i, 1 < i < total\ band, i \in \mathbb{Z}}) \right]$$

where $f(k)$ is the fluctuation ratio, $\{M(k)\}_{k \in \mathbb{N}}$ is the MEE sequence of $k$ beats from one recording. $M_{ref}(k)$ is the reference metric, $Mid$ denote the median amplitude value of the corresponding bandwidths.

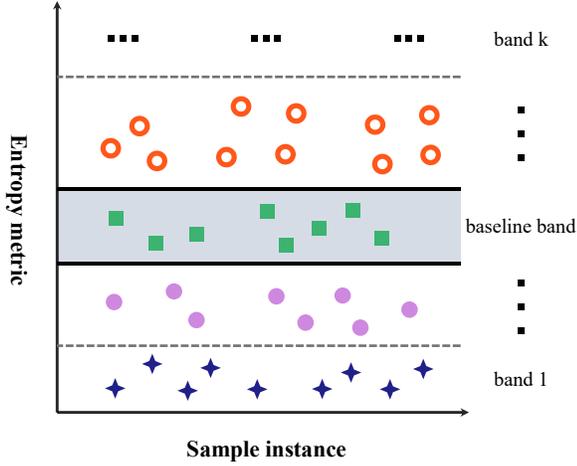

**Fig. 2.** Illustration of the proposed bandwidth picking method. The bandwidth range with the maximum number of samples is identified as the reference to compute the degree of fluctuation of other samples relative to the reference.

Since the threshold is artificially determined, the subsequent assessment can also be graded according to the fluctuation ratio $f(k)$. And as $f(k)$ increases, the confidence level that an abnormal heartbeat is judged increases accordingly.

Based on the proposed bandwidth picking method, MEE can also be utilized for the evaluation of sample diversity. For samples with the same label from different subjects, the data set is finally cropped by counting the number of samples appearing in different bandwidths and discarding the bandwidths with multiple samples, which can eliminate redundant data and shorten the time of model training.

## 4. APPLICATIONS AND EXPERIMENTAL RESULTS

### 4.1. Applications of Proposed MEE

The aim of this work is to propose a measure of ECG morphological chaos that is well adapted to ECG morphology. Considering the inter-patient variability [18, 19], different subjects have a diversity of ECG morphological manifestations for arrhythmia, so ECG signals with different label from different subjects may have similar metric values. The MEE metric is divided into two application modes: Mode I for arrhythmia screening from individual patient ECG recordings (assuming that the arrhythmia signal has a significant difference from the normal signal) and Mode II for diversity evaluation of the same arrhythmia between different subjects, this mode is utilized to measure the degree of difference between subjects, which helps to provide richer feature information for the subsequent construction of classification model.

Based on the above two application modes, we conduct application experiments corresponding to the above two modes, which are application I-Unsupervised rapid screening of arrhythmias, II-Representative data selection for dataset pruning and establishment of diverse dataset. To explore more application scenarios, we also conduct the following application explorations: application III-Quality assessment, and design a clinical interactive interface based on the proposed MEE metric.

In the existing researches, the data sets are divided in a random way without considering the diversity of the training samples, which is due to the lack of a measure of sample diversity. Based on the MEE metric designed in this paper, it can guide the training data grouping and expand the feature space of single class information as much as possible, providing an interpretable way to group training data. Moreover, the MEE metric offers another possibility based on its ability of capturing areas with significant morphological differences. Unlike the arrhythmia screening of application I, the morphological changes between different arrhythmias are in a small range due to the abnormal heartbeats often correspond to morphological changes in P waves, QRS complexes, T waves, etc. [20, 21], so their morphological changes are often in a controlled interval. However, for signal segments with poor signal quality, their morphology often does not possess any pattern with structure, and thus their corresponding MEE values should be significantly different and sharply unstable compared to signals with good signal quality. The above assumptions will be utilized to explore an automated label-free method for the detection of signal segments with poor signal quality based on MEE metric.

In the subsequent exploration of MEE indicators, we mainly focus on three experimental aspects: beat-level-based arrhythmia screening, diversity optimization of heartbeat-single lead-standard 12-lead ECG, and noise region detection. These experiments demonstrate the good scalability of MEE. We employ the classical Pan-Tompkins (PT) algorithm [22] to locate the R-peak and refine its position by searching for extreme values in nearby regions. Using the reference coordinates of the R-peak, a single beat is extracted by intercepting 0.4 seconds before and after it, for subsequent evaluation. Moreover, since both BWE and WSE require a bandwidth parameter, a value of 100 is employed in this study. This implies that the normalized amplitude range is partitioned into 100 equal segments for further computations.

### 4.2. Experimental Details – Application I: Unsupervised Rapid Screening of Arrhythmias

To validate the effectiveness of MEE for arrhythmia screening of single recordings, we utilize a classic arrhythmia

**Table 1**
EXPERIMENTS OF APPLICATION I: BEAT-LEVEL ARRHYTHMIA SCREENING

| Metric | PE | | | AE | | | SE | | | FE | | | MEE-I | | | MEE-II | | | MEE-III | | | MEE-IV | | |
|---|---|---|---|---|---|---|---|---|---|---|---|---|---|---|---|---|---|---|---|---|---|---|---|---|
| Evaluation | ACC | SPE | F1 | ACC | SPE | F1 | ACC | SPE | F1 | ACC | SPE | F1 | ACC | SPE | F1 | ACC | SPE | F1 | ACC | SPE | F1 | ACC | SPE | F1 |
| #106 | 0.755 | 0.824 | 0.569 | 0.281 | 0.000 | 0.438 | 0.651 | 0.635 | 0.527 | 0.831 | 0.911 | 0.675 | **0.891** | 0.974 | **0.776** | 0.885 | 0.963 | 0.771 | 0.879 | **0.978** | 0.742 | 0.876 | 0.968 | 0.744 |
| #119 | 0.630 | 0.594 | 0.509 | 0.359 | 0.160 | 0.429 | 0.770 | 0.828 | 0.581 | 0.842 | **0.958** | 0.632 | 0.886 | 0.937 | 0.774 | **0.918** | 0.947 | **0.842** | 0.694 | 0.724 | 0.512 | 0.878 | 0.932 | 0.758 |
| #200 | 0.597 | 0.519 | 0.572 | 0.369 | 0.000 | 0.539 | 0.369 | 0.000 | 0.539 | 0.712 | 0.626 | 0.688 | **0.947** | 0.990 | **0.925** | 0.946 | 0.988 | 0.923 | 0.367 | 0.000 | 0.537 | 0.945 | 0.988 | 0.921 |
| #201 | 0.837 | 0.909 | 0.419 | 0.877 | 0.960 | 0.473 | 0.767 | 0.825 | 0.341 | 0.967 | 0.995 | 0.876 | 0.974 | **0.998** | 0.901 | 0.974 | 0.998 | 0.903 | 0.974 | 0.998 | 0.903 | **0.975** | 0.998 | **0.905** |
| #208 | 0.478 | 0.000 | 0.646 | 0.478 | 0.000 | 0.646 | 0.478 | 0.000 | 0.646 | 0.630 | 0.487 | 0.669 | 0.685 | 0.659 | 0.684 | **0.743** | **0.785** | **0.721** | 0.478 | 0.000 | 0.646 | 0.520 | 0.137 | 0.651 |
| #210 | 0.811 | 0.857 | 0.249 | 0.922 | 0.971 | 0.496 | 0.910 | 0.945 | 0.522 | 0.941 | 0.993 | 0.562 | 0.977 | 0.991 | 0.865 | **0.978** | 0.992 | **0.873** | 0.977 | **0.995** | 0.863 | 0.977 | 0.991 | 0.865 |
| #219 | 0.478 | 0.000 | 0.646 | 0.478 | 0.000 | 0.646 | 0.478 | 0.000 | 0.646 | 0.630 | 0.487 | 0.669 | 0.685 | 0.659 | 0.684 | **0.743** | **0.785** | **0.721** | 0.478 | 0.000 | 0.646 | 0.520 | 0.137 | 0.651 |
| #221 | 0.837 | 0.909 | 0.419 | 0.877 | 0.960 | 0.473 | 0.767 | 0.825 | 0.341 | 0.967 | 0.995 | 0.876 | 0.974 | **0.998** | 0.901 | 0.974 | 0.998 | 0.903 | 0.974 | 0.998 | 0.903 | **0.975** | 0.998 | **0.905** |
| #223 | 0.204 | 0.000 | 0.339 | 0.275 | 0.101 | 0.349 | 0.854 | 0.925 | 0.618 | 0.903 | 0.990 | 0.704 | 0.924 | 0.995 | 0.777 | **0.931** | **0.996** | **0.800** | 0.919 | 0.983 | 0.771 | 0.928 | 0.995 | 0.792 |
| #233 | 0.468 | 0.329 | 0.468 | 0.600 | 0.583 | 0.482 | 0.291 | 0.000 | 0.451 | **0.853** | **0.942** | 0.715 | 0.843 | 0.793 | 0.782 | 0.841 | 0.790 | 0.779 | 0.846 | 0.794 | **0.786** | 0.846 | 0.795 | 0.786 |
| #Total | 0.610 | 0.494 | 0.484 | 0.552 | 0.374 | 0.497 | 0.634 | 0.498 | 0.521 | 0.828 | 0.838 | 0.707 | 0.879 | 0.899 | 0.807 | **0.893** | **0.924** | **0.824** | 0.759 | 0.647 | 0.731 | 0.844 | 0.794 | 0.798 |

*The best performance is demonstrated in screening each patient, contrasting the potential for differentiation among various indicators. MEE-I: WSE-I+ FT-I. MEE-II: WSE-II+FT-I. MEE-III: WSE-I+FT-II. MEE-I: WSE-II+FT-II. WSE-I represents preserving the amplitude of the wave sets below the baseline bandwidth and utilizing the amplitudes that are all negative for entropy calculation. WSE-II represents reversing the amplitude of the wave sets below the baseline bandwidth and utilizing only positive amplitude for entropy calculation. FT-I and FT-II respectively represent the MEE fusion types based on mean square value $(BWE^2+WSE^2)/2$ and improved Euler's formula. Beat distribution details: *#106*→N:1505, S:0, V:518, F:0, Q:0, O:69. *#119*→N:1539, S:0, V:443, F:0, Q:0, O:106. *#200*→N:1739, S:30, V:825, F:2, Q:0, O:190. *#201*→N:1631, S:128, V:198, F:2, Q:0, O:74. *#208*→N:1584, S:2, V:990, F:372, Q:2, O:84. *#210*→N:2419, S:22, V:194, F:10, Q:0, O:34. *#219*→N:2077, S:7, V:64, F:1, Q:0, O:157. *#221*→N:2026, S:0, V:396, F:0, Q:0, O:34. *#223*→N:2040, S:73, V:473, F:14, Q:0, O:37. *#233*→N:2226, S:7, V:830, F:11, Q:0, O:72. *#Total*→N:18786, S:269, V:4931, F:412, Q:2, O:857.

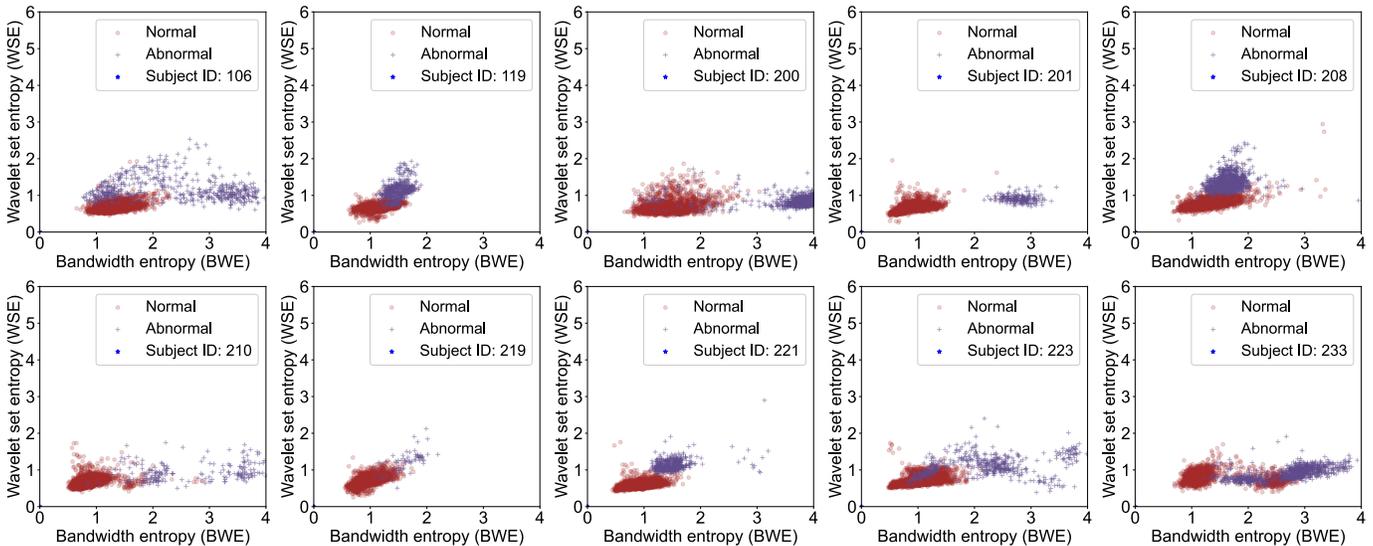

**Fig. 3.** The performance of BWE and WSE metrics across different patients is illustrated through scatter plots, where red and purple indicate normal or abnormal beat-level samples, respectively.

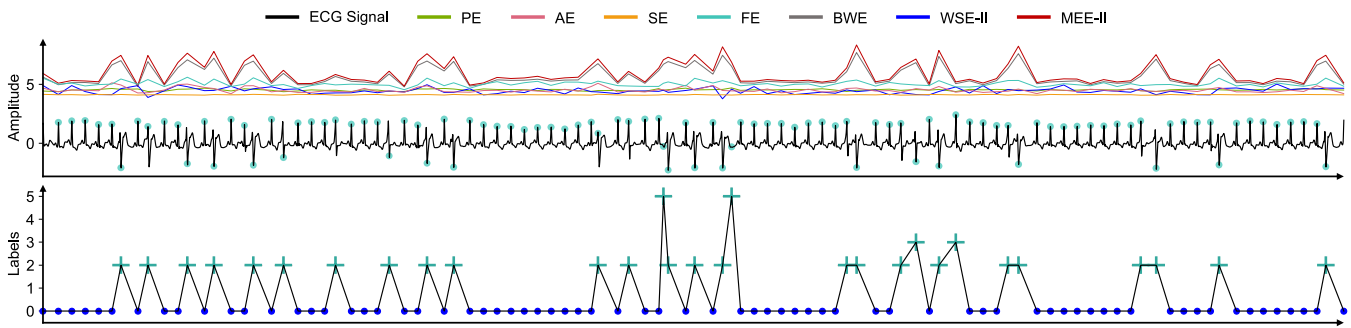

**Fig. 4.** The evaluation results of MEE corresponding to 100-beat lengths are visualized (Patient ID 233, lead II), demonstrating consistency with the annotations of true beat labels, with pronounced deviations in abnormal beat positions compared to the surrounding abnormal fluctuation values. A comparison of different metrics showcases the superiority of MEE-II. Label 0 represents normal heartbeats, while labels 1-5 represent SVEB, VEB, fusion beat, unclassified beat, and other types. The values of all metrics have been shifted upwards by 4 to ensure no overlap with the original signal.

databases which have detailed annotation information for individual heartbeats and allow a more precise assessment of the applicability of MEE. The database and experiments are described as follows:

*1) MIT-BIH Arrhythmia Database (MIT-DB):* MIT-DB is widely applied in ECG classification for algorithm evaluation [23]. Each recording in MIT-DB consists of two leads with a duration of 30 min and a sampling frequency of 360 Hz. The first lead corresponds to Lead II, while the second lead corresponds to Lead V1 and sometimes to V2, V4, or V5. This database contains 48 recordings, of which (102, 104, 107, and 217) are normally removed because they contain paced heartbeats.

   *2) Experiments:* To evaluate the proposed MEE, we select patients from the MIT-DB with morphological abnormalities. These patients exhibit sufficient ventricular ectopic beats (VEBs) and include various morphological abnormalities such as fusion beats, unclassified beats, and other types. These types exhibit distinguishable morphological differences from sinus beats (SBs). Considering these factors, we combine all these abnormal types and measure the ability of MEE to differentiate them from SBs. Given the need for an adequate variety of morphological abnormal heartbeat types, we select patients with the following IDs: 106, 119, 200, 201, 208, 210, 219, 221, 223, and 233. We perform detailed estimations on all consecutive heartbeats of these 10 patients. Additionally, it is noteworthy that supraventricular ectopic beats (SVEBs) are not considered in this study as the focus is on morphological information assessment. SVEBs require diagnosis based on rhythm information, and their individual heartbeat analysis renders them indistinguishable from SBs morphologically. We will discuss additional settings for SVEBs detection using MEE in the discussion section of this paper.

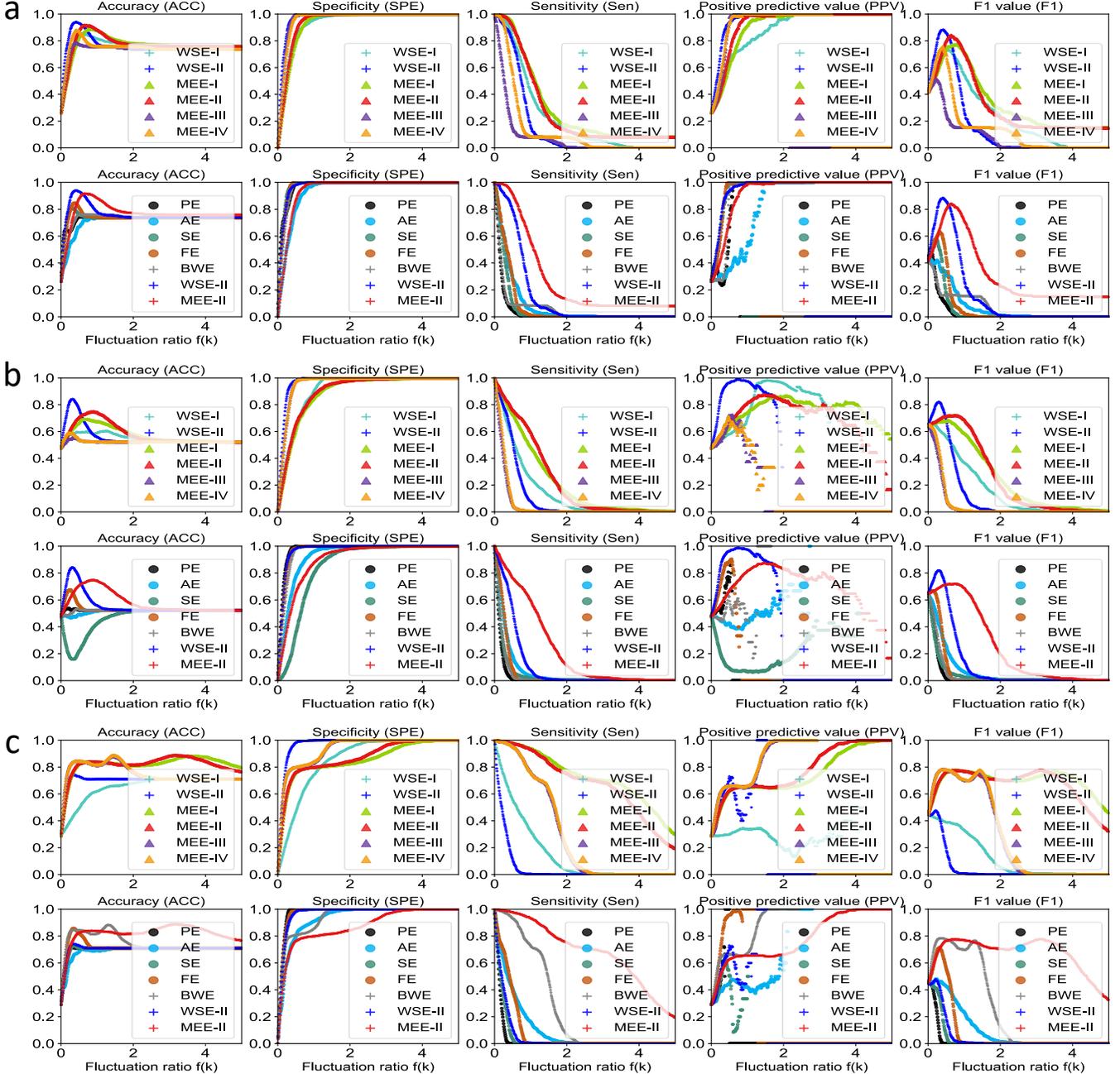

**Fig. 5.** Different entropy metrics and configurations (details are provided below in Table 1) demonstrate potential performance in patient-level arrhythmia screening. (a)-(c), corresponding to patients 119, 208, and 233, are evaluated using lead II. Through a grid search of the fluctuation ratio f(k), the metrics' ability to detect arrhythmias at the beat level is thoroughly investigated, including optimal performance and stability in response to changes in f(k).

On the testing records of 10 patients (Table 1), the MEE calculation based on Type II combination achieves the highest accuracy of 0.893, specificity of 0.924, and F1 score of 0.824, exhibiting the greatest potential for arrhythmia screening among the four combinations and other entropy-based metrics.

To demonstrate the discriminative ability of the proposed BWE and WSE for morphological anomalies in a more intuitive manner, we have plotted scatter diagrams for the 10 tested patients, with BWE representing the amplitude information on the horizontal axis and WSE representing the phase information on the vertical axis. This visualization allows for a visual inspection of the potential discriminative power of the different metrics across individuals (Fig. 3).

From the scatter plot, we observe that the detection of abnormal morphology features and normal morphology features shows better discrimination when the BWE values of abnormal morphology heartbeats are concentrated in the region greater than 2-3. The BWE value of the abnormal morphology heartbeats sample population corresponds to an inverted heartbeat waveform in the region greater than 2-3, which is the reason for the significant high value of BWE. This further proves the sensitive characteristics of BWE to inverted abnormal morphology. However, for the case where the BWE value of the abnormal morphology heartbeats sample population is concentrated in the region less than 2, the corresponding abnormal morphology waveform is generally positive (i.e., the R-peak in the QRS complex is above), and at this point, WSE shows a more significant distinguishable characteristic than BWE. This is because the amplitude domain entropy of positive abnormal morphology heartbeats and normal heartbeats is relatively close, making it difficult to distinguish them. The distinguishable characteristic of WSE in this case demonstrates the feasibility of utilizing entropy measures based on phase information to test abnormal morphology. A more intuitive example provides a good illustration of the hypothesis. It is observed from the scatter plot of the patient ID 106 that BWE has a more significant discriminative power when BWE is greater than 2-3, whereas WSE exhibits a more significant discriminative power when it is less than 2-3. This is the first observation of the hypothesis from a single subject, confirming that BWE and WSE have significant discriminative power even when both normal and abnormal waveforms exist in the same patient, regardless of their directionality. Additionally, we present screening examples of 100 heartbeats for patient 233 (Fig. 4), where we compare the fluctuation rates of PE, AE, SE, FE indices, as well as the BWE, WSE-II, and MEE-II proposed in this paper, with the visual consistency of human perception at morphologically abnormal heartbeats. Each heartbeat is labeled with an abnormal tag, and it can be intuitively observed that the unsupervised measurement based on MEE-II demonstrates good consistency with abnormal morphological variations.

Based on the scatter plot results in Fig. 3, we summarize three screening patterns: 1). Both BWE and WSE have the ability to distinguish abnormal morphology. 2). WSE can distinguish while BWE cannot. 3). BWE can distinguish while WSE cannot. To demonstrate the superiority of MEE, we perform a grid

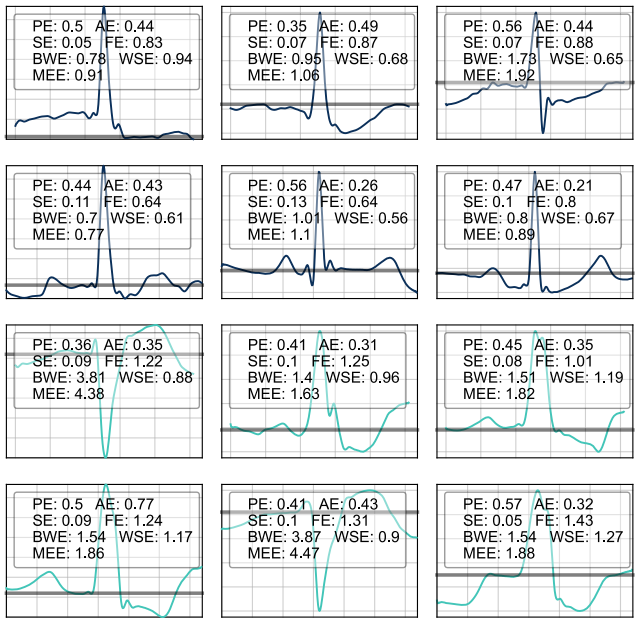

**Fig. 6.** The diversity analysis of beat-level samples is demonstrated, with the first two rows representing normal sinus beats and the latter two rows representing premature ventricular contractions (PVCs). The black horizontal line represents the determined approximate baseline. A comparison is made regarding the ability of MEE and other metrics to reflect morphological differences among samples within the same category, aiding in the assessment of diversity among similar samples at the beat level.

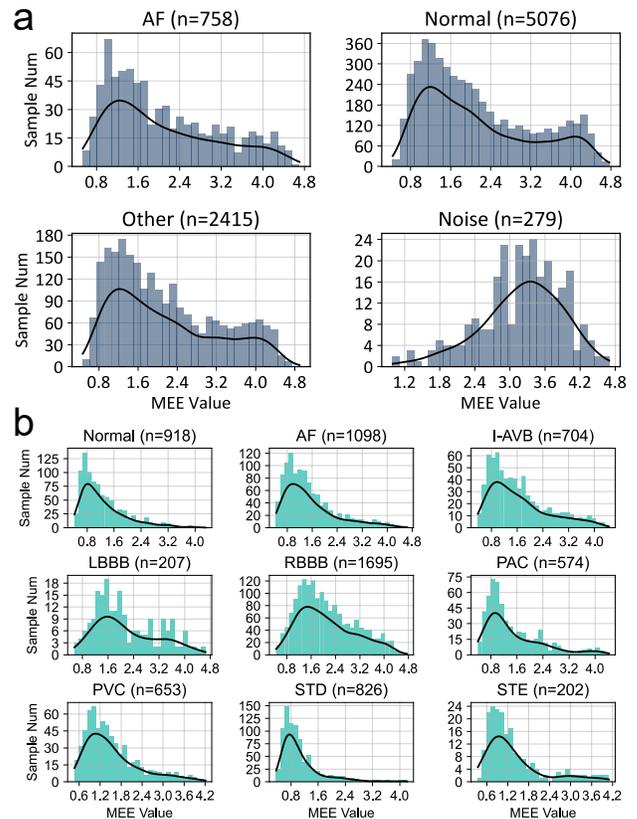

**Fig. 7.** Using beat-level computation, MEE values are assigned to each record, and a histogram distribution is generated to count the number of records within different MEE ranges, while simultaneously presenting a kernel density estimation curve. (a) 2017 PhysioNet/CinC Database. (b) 2018 China Physiological Signal Challenge (CPSC) Database.

**Table 2**
DESCRIPTION OF UTILIZED DATABASES

| Database (Sampling rate) | Lead type | Label type | Number of recordings (Train/Test) | Signal length (sample-point) | | | |
|---|---|---|---|---|---|---|---|
| | | | | Max | Min | Average | Std |
| 2017 PhysioNet/CinC Database (300 Hz) | Single-lead | Normal sinus rhythm (Normal) | 5076/150 | 18286/18170 | 2714/3002 | 9633/9708 | 2992/3280 |
| | | Atrial fibrillation (AF) | 758/50 | 18062/18000 | 2996/3020 | 9703/9133 | 3695/3025 |
| | | Alternative rhythm (Other) | 2415/70 | 18258/18000 | 2738/3064 | 10289/7549 | 3529/3084 |
| | | Noise | 279/30 | 18000/18000 | 2808/3064 | 7314/7549 | 3122/3084 |
| | | Total | 8528/300 | 18286/18170 | 2714/3002 | 9749/9636 | 3267/3444 |
| 2018 China Physiological Signal Challenge (CPSC) Database (500 Hz) | Standard 12-lead | Normal | 918/45 | 32000/32000 | 5000/5000 | 7714/8844 | 3816/5288 |
| | | AF | 1098/43 | 37000/22498 | 4500/5000 | 7521/7942 | 4300/4647 |
| | | First-degree atrioventricular block (I-AVB) | 704/25 | 27157/14000 | 4999/4999 | 7133/6517 | 3501/2164 |
| | | Left bundle branch block (LBBB) | 207/11 | 32500/25179 | 4500/5000 | 7472/8878 | 4106/6286 |
| | | Right bundle branch block (RBBB) | 1695/71 | 59000/21352 | 4999/5000 | 7308/7153 | 4350/3705 |
| | | Premature atrial contraction (PAC) | 556/26 | 37000/30000 | 4500/5000 | 9717/9406 | 6223/6719 |
| | | Premature ventricular contraction (PVC) | 672/30 | 72000/28556 | 3000/5000 | 10459/9181 | 7616/5599 |
| | | ST-segment depression (STD) | 825/33 | 69000/15500 | 4000/5000 | 7752/7710 | 4826/2713 |
| | | ST-segment elevated (STE) | 202/16 | 30000/23050 | 5000/5000 | 8574/9129 | 5349/5317 |
| | | Total | 6877/300 | 72000/32000 | 3000/4999 | 7974/8095 | 4979/4755 |

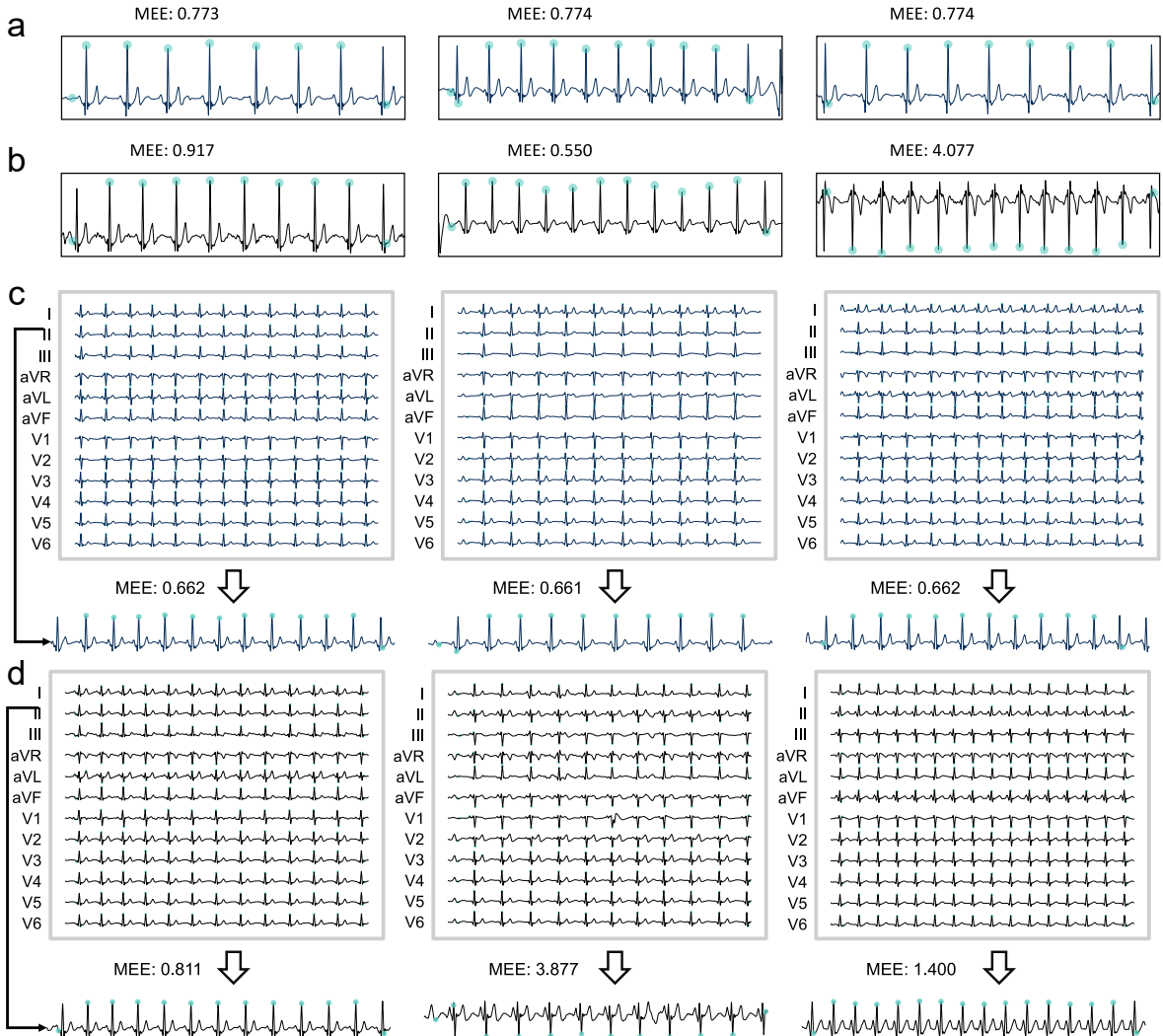

**Fig. 8.** Visualizing the diversity of single-lead and 12-lead ECGs. Diversity control groups are designed, consisting of three samples with close MEE computed values and three samples randomly selected with differing MEE values. Guidance optimization for 12-lead ECG is based on lead II. (a) 2017 PhysioNet/CinC Database. (b) 2018 China Physiological Signal Challenge (CPSC) Database.

search on the fluctuation rate threshold, ranging from 0 to 5 with a step size of 0.01, and thoroughly explore comparisons with existing classical methods (PE, AE, SE, FE) as well as different combinations of our proposed methods (WSE-I, WSE-II, MEE-I~IV). The three patterns are respectively applied to representative patients with IDs 119, 208, 233. The results are shown in Fig. 5. We find that the MEE-II type exhibits the best overall performance among the three patterns and significantly outperforms traditional methods. Additionally, we identify an outstanding property of MEE-II, which is its robustness to changes in fluctuation rate, avoiding sudden performance

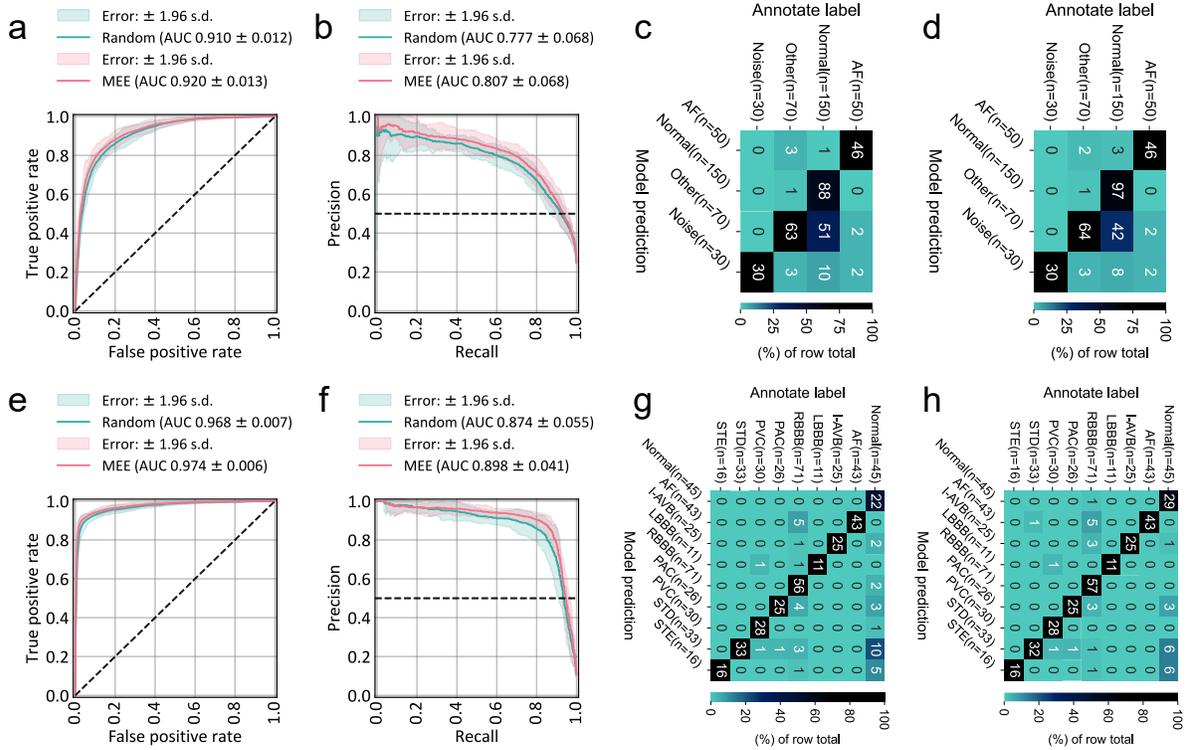

**Fig. 9.** Different data pruning methods (random pruning and MEE-based pruning) are compared in terms of their corresponding classification results. The model training is repeated 10 times, providing means and standard deviations to mitigate randomness. (a)-(d) represent the ROC and PRC curves tested on the 2017 PhysioNet/CinC Database, as well as the confusion matrices for random pruning models and MEE-based pruning models with median performance. The ROC and PRC curves in (a)-(d) provide means and 95% confidence intervals. (e)-(h) represent the evaluation results on the 2018 China Physiological Signal Challenge (CPSC) Database.

Table 3

PERFORMANCE EVALUATION OF DIVERSITY OPTIMIZATION IN SINGLE-LEAD AND TWELVE-LEAD CONFIGURATIONS USING DIFFERENT PRUNING METHODS

| Dataset | Category | Random | | | | | MEE | | | | |
|---|---|---|---|---|---|---|---|---|---|---|---|
| | | SEN | SPE | PPV | F1 | ACC | SEN | SPE | PPV | F1 | ACC |
| 2017 PhysioNet/CinC Database | AF | 0.904±0.043 | 0.975±0.015 | 0.883±0.056 | 0.891±0.030 | 0.755 ±0.023 | **0.914±0.025** | **0.978±0.015** | **0.895±0.055** | **0.903±0.030** | **0.788 ±0.025** |
| | Normal | 0.594±0.065 | **0.983±0.012** | 0.973±0.015 | 0.735±0.046 | | **0.659±0.064** | 0.981±0.019 | **0.974±0.022** | **0.784±0.039** | |
| | Other | **0.893±0.028** | 0.777±0.043 | 0.553±0.042 | 0.682±0.024 | | 0.884±0.041 | **0.816±0.033** | **0.597±0.036** | **0.711±0.018** | |
| | Noise | 0.990±0.015 | 0.951±0.008 | 0.693±0.037 | 0.814±0.027 | | **1.000±0.000** | **0.953±0.004** | **0.705±0.017** | **0.827±0.012** | |
| 2018 China Physiological Signal Challenge (CPSC) Database | Normal | 0.569±0.068 | **0.996±0.004** | 0.959±0.043 | 0.712±0.059 | 0.860 ±0.016 | **0.647±0.044** | 0.996±0.003 | **0.971±0.023** | **0.775±0.030** | **0.888 ±0.009** |
| | AF | 0.998±0.007 | 0.977±0.005 | 0.878±0.024 | 0.934±0.013 | | **1.000±0.000** | **0.983±0.004** | **0.908±0.019** | **0.951±0.011** | |
| | I-AVB | 0.988±0.026 | 0.985±0.007 | 0.862±0.059 | 0.919±0.036 | | **1.000±0.000** | 0.985±0.007 | 0.859±0.053 | **0.924±0.031** | |
| | LBBB | **1.000±0.000** | **0.996±0.001** | 0.903±0.028 | 0.949±0.016 | | **1.000±0.000** | 0.996±0.003 | **0.907±0.072** | **0.950±0.040** | |
| | RBBB | 0.742±0.040 | **0.996±0.003** | **0.982±0.014** | 0.845±0.027 | | **0.789±0.041** | 0.995±0.004 | 0.981±0.014 | **0.874±0.023** | |
| | PAC | 0.965±0.021 | 0.973±0.009 | 0.774±0.057 | 0.858±0.036 | | **0.973±0.025** | **0.979±0.008** | **0.821±0.059** | **0.889±0.029** | |
| | PVC | **0.947±0.022** | 0.996±0.005 | 0.961±0.042 | 0.953±0.024 | | **0.947±0.027** | **0.999±0.002** | **0.990±0.016** | **0.967±0.016** | |
| | STD | 0.967±0.034 | 0.962±0.010 | 0.761±0.046 | 0.850±0.028 | | **0.988±0.015** | **0.973±0.009** | **0.822±0.045** | **0.896±0.025** | |
| | STE | 0.988±0.025 | 0.966±0.014 | 0.635±0.087 | 0.770±0.071 | | **0.994±0.019** | **0.969±0.011** | **0.658±0.087** | **0.788±0.058** | |

*A ± B: Mean ± standard deviation.

degradation. This provides ample tolerance space for selecting fluctuation rates in screening.

Moreover, this further demonstrates that when the BWE range of anomalous morphology heartbeat clusters are below 2-3 (i.e., positive QRS complex), WSE exhibits superior anomalous morphology detection characteristics. Therefore, it is necessary to appropriately increase the contribution weight of WSE to MEE for positive anomalous heartbeats. The above analysis thoroughly demonstrates the necessity of combining corresponding amplitude and phase information from BWE and WSE, which can provide a more comprehensive reference for abnormal morphology detection.

### 4.3. Experimental Details – Application II: Representative Data Selection for Dataset Pruning and Establishment of Diverse Dataset

To further explore the additional value of MEE, we design sample diversity optimization experiments. The dataset and experimental descriptions are as follows:

***1) 2017 PhysioNet/CinC Database & 2018 China Physiological Signal Challenge (CPSC) Database:*** The 2017 PhysioNet/CinC challenge dataset [24] aim to classify single-lead ECG recordings to four types: normal sinus rhythm (Normal), atrial fibrillation (AF), alternative rhythm (Other) or noise (Noise), which contain 8,528 single-lead ECG recordings lasting from 9 to 60 seconds with a sampling rate of 300 Hz, including 5,076 Normal, 758 AF, 2,415 Other and 279 Noise examples.

The CPSC2018 dataset [25] includes 12-lead ECG recordings from 6,877 patients, comprising 3,178 females and 3,699 males. These ECG recordings are acquired from 11 hospitals. The length of the recordings lengths varying from 6 to 60 seconds, and the sampling rate is 500 Hz. The challenge is aim to classify 12-lead ECG recordings to nine types: Normal, AF, First-degree atrioventricular block (I-AVB), Left bundle branch block (LBBB), Right bundle branch block (RBBB), Premature atrial contraction (PAC), Premature ventricular contraction (PVC), ST-segment depression (STD), and ST-segment elevated (STE). More details are shown in Table 2.

*2) Experiments:* The proposed MEE has the potential to evaluate the diversity of samples within the same category. We conduct diversity analysis on Physionet2017 and CPSC2018 databases. It is worth noting that the screening of arrhythmias (Application I) is performed on different arrhythmia categories of individual subjects, while diversity analysis is conducted on the same category of arrhythmias from different patients. To demonstrate the potential application of the proposed morphological metric in evaluating diversity, we perform dataset pruning by removing redundant samples of the same category (samples with MEE values within a small error range are considered redundant). We compare the pruned dataset with randomly pruned samples to demonstrate the value of the proposed method (random pruning leads to loss of feature space and affects the subsequent classification performance of DL models [26]). We use a 1D ResNet50 network [27, 28] for classification, utilizing test data drawn from the original publicly released dataset consisting of 300 samples, encompassing various arrhythmia categories. We fix the model weights based on the best performance on the validation set to avoid randomness and train the model 10 times to average the performance. Fig. 6 demonstrates the diversity assessment at beat-level, showcasing strong capabilities in diversity assessment compared to existing classical metrics. Utilizing beat-level MEE, we compute the average MEE value for the entire lead II record to obtain a record-level evaluation. To alleviate edge effects, ECG recordings exclude heartbeats at both ends from MEE calculation. Fig. 7 illustrates the distribution of MEE for different categories, indicating that the majority of samples are concentrated in lower MEE ranges, which represent areas of redundant information. Concurrently, we visualize the morphological comparison of redundant samples between single-lead and standard 12-lead records (Fig. 8), as well as showcase diverse samples with noticeable differences in MEE. Based on these visualization results, our focus primarily lies on pruning normal signals, as the number of abnormal heartbeats in abnormal signals varies, posing a continued challenge. Moreover, in real-world scenarios, normal samples vastly outnumber positive samples; hence, pruning normal samples to match the number of positive samples mitigates class imbalance issues and is deemed necessary. We invite three experts in cardiac electrophysiology to evaluate the consistency of MEE metric and morphological differences and received their approval. The advantage of this method in diversity evaluation will contribute to the construction of diverse databases, preserve more valuable samples under the trend of big data, and greatly reduce the redundancy of data storage.

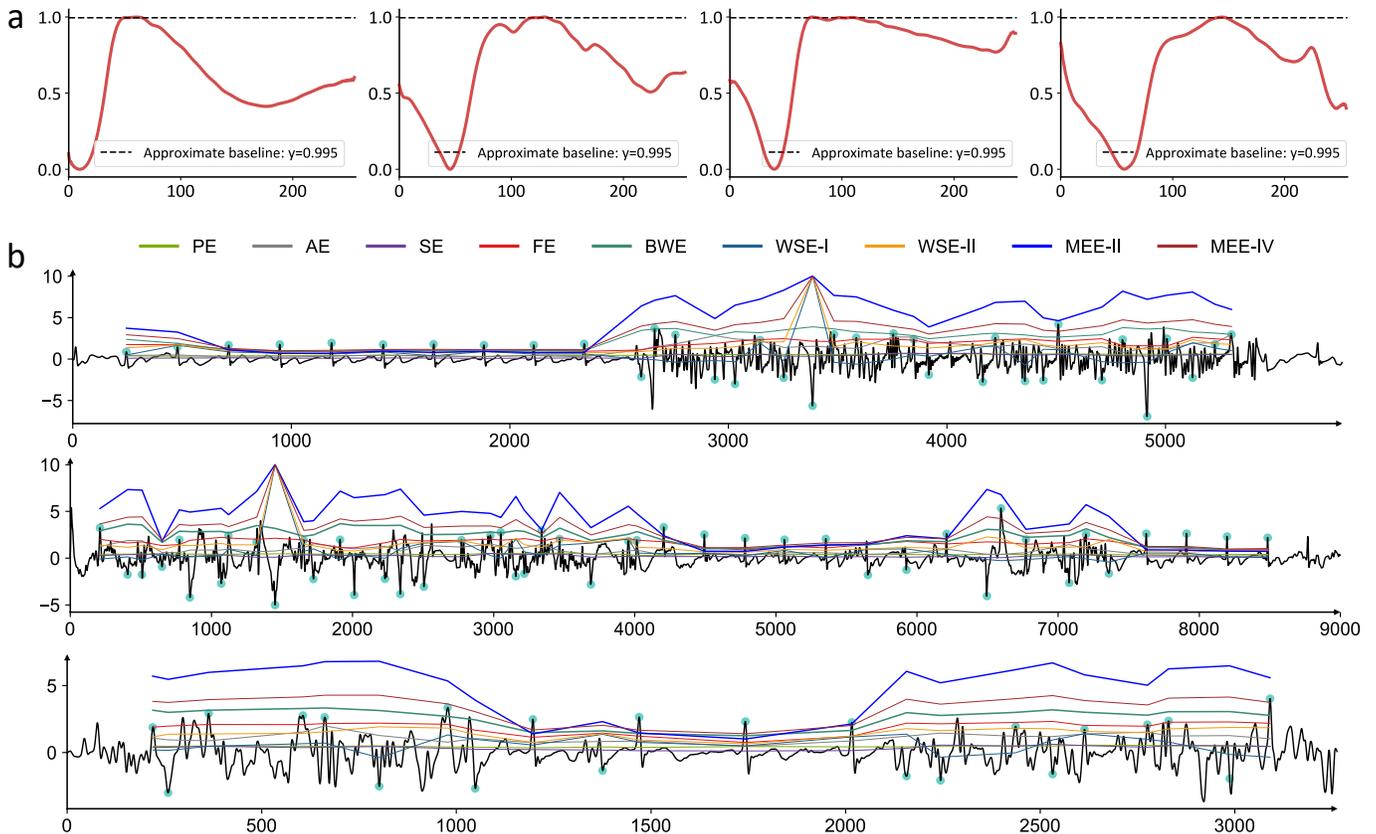

**Fig. 10.** The ability of MEE to localize noise-related morphologies is demonstrated. (a) The identification results of the approximate baseline for morphologically abnormal segments produced by WSE. (b) The recognition results of three noise segments from Physionet2017 are presented, comparing different entropy metrics.

We separately evaluate the performance comparison between random pruning and MEE-guided pruning based on the two databases (Fig. 9 and Table 3). The results demonstrate that MEE-guided pruning significantly outperforms random pruning in classification performance, with overall accuracies improving from 0.755±0.023 to 0.788±0.025 and from 0.860±0.016 to 0.888±0.009, respectively. The overall AUROC values also increase from 0.910±0.012 to 0.920±0.013 and from 0.968±0.007 to 0.974±0.006, respectively. Additionally, the AUPRC values increase from 0.777±0.068 to 0.807±0.068 and from 0.874±0.055 to 0.898±0.041, respectively. This further confirms that random pruning discards useful feature representations, leading to a decrease in representational richness and weakening the model's ability to discriminate between different category boundaries.

### 4.4. Experimental results – Application III: Signal Quality Assessment

We also present the capability of MEE to identify noise regions. WSE is more sensitive to noise segments when determining the approximate baseline. Fig. 10a illustrates four examples where the approximate baseline is determined as the topmost bandwidth, resulting in all wavelet sets being on the same side, leading to abnormal WSE values. Through this property, noise localization can also be achieved. In long-duration records, due to the absence of inherent morphological patterns in noise segments, larger abnormal values may be obtained. To demonstrate this, we utilize samples labeled as Noise in the Physionet2017 database and randomly select three examples to compare the recognition capabilities of different entropy methods for noise regions (Fig. 10b). We use the PT [22] algorithm to locate the R-peak positions and extract both ends of the R-peak recognition results to avoid edge effects. Experienced electrocardiologists are invited to evaluate the ability of MEE to detect noise through morphology. Results indicate that MEE, as a new ECG morphological entropy index, exhibits outstanding detection capabilities for both ECG arrhythmic morphological abnormalities and noise morphological abnormalities, demonstrating good visual consistency and outperforming existing classical metrics.

## 5. DISCUSSION

As a non-stationary system, previous researchers have utilized methods such as AE [29, 30], SE [10, 31], FE [32, 33], and PE [6, 34] to analyze ECG signals. AE, SE, and FE, as a series of derivative methods, primarily focus on statistically repeating patterns in time series and form metrics to measure irregularity and randomness within the sequence. PE, on the other hand, evaluates the regularity and periodicity of sequences by exploring sorting patterns under specified orders in the time series, aiding in screening for cardiac arrhythmias. However, AE, SE, and PE have lower tolerance for noise. Common Gaussian noise in the real world introduces randomness into the sequence, potentially masking patterns within the sequence and reducing the ability of PE to capture the true structure of the sequence. FE forms a more robust indicator for noise by considering the fuzziness and uncertainty in the signal. To demonstrate the robustness of different methods in heartbeat measurement against noise, we compare the introduction of Gaussian distribution noise with different standard deviations 0.01, 0.02, and 0.03 (Fig. 11). The results indicate that FE, along with the three metrics proposed by us (BWE, WSE, MEE), demonstrates good robustness against noise. This is because BWE calculates entropy values based on bandwidth, and the introduction of Gaussian noise does not significantly disrupt the pattern of bandwidth distribution. Similarly, WSE calculates energy based on wavelet sets, and the introduction of noise does not greatly disturb the magnitude of energy. Therefore, both experimental and analytical results demonstrate the excellent robustness of MEE and FE against noise interference.

**Table 4**
COMPUTATIONAL EFFICIENCY COMPARISON OF DIFFERENT METRICS

| Method | PE | AE | SE | FE | BWE | WSE-II | MEE-I | MEE-II |
|---|---|---|---|---|---|---|---|---|
| Detail | order = 4, step size = 1 | m = 2, r = 0.2×s.d. | m = 2, r = 0.2×s.d. | m = 2, r = 0.2×s.d., n=2 | bandwidth = 100 | bandwidth = 100 | $\frac{BWE^2 + WSE^2}{2}$ | $BWE \times e^{\frac{WSE}{2\pi}}$ |
| Processing time (ms) | 8.3951 | 1273.6 | 1159.7 | 1177.6 | 1.2155 | 2.8979 | 0.0003 | 0.0003 |

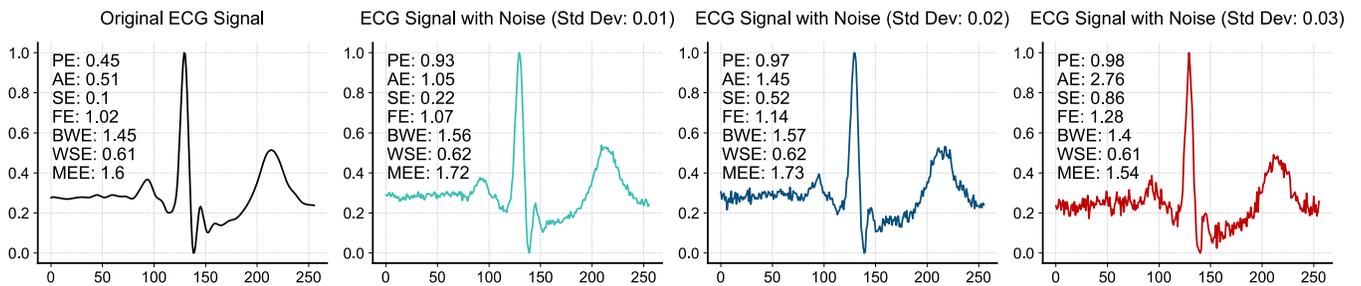

**Fig. 11.** An example demonstrating the robustness of different entropy metrics to noise.

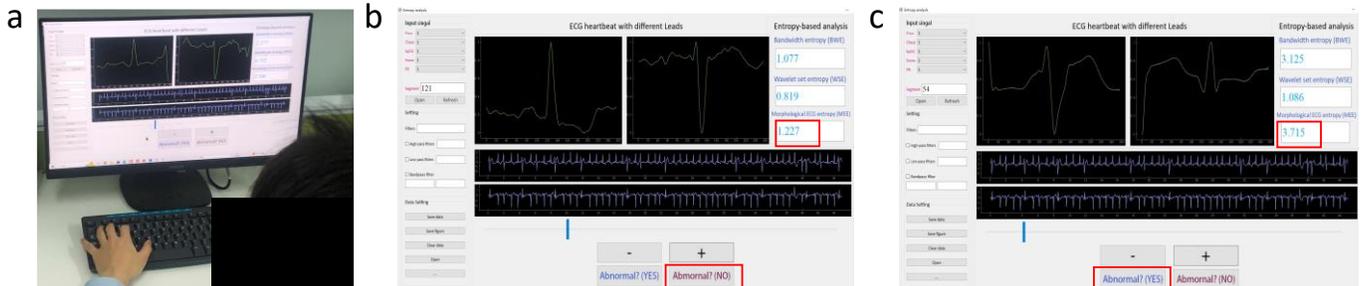

**Fig. 12.** Clinical Work Interface and Detailed Information. (a) Operator interface physical diagram. (b) Sinus beat sample with moderate MEE value. (c) Ventricular premature beat sample with abrupt MEE value.

Considering the need for real-time monitoring, we compare the average computation time per heartbeat for different entropy methods, all tested on an i5-12600KF CPU (Table 4). The results demonstrate a significant reduction in computation time for MEE compared to AE, SE, and FE, by approximately 400 times. Given that a single heartbeat lasts approximately 1-second, AE, SE, and FE are difficult to achieve real-time monitoring without special processing. The study highlights the efficiency of the MEE metric, with computation time per heartbeat averaging around 4ms, making it suitable for real-time applications. This rapid computational capability enables the swift screening of large volumes of ECG data, thereby alleviating the burden on doctors and increasing feasibility in clinical practice.

Furthermore, to delve deeper into the nature of the proposed MEE, scatter plots of BWE and WSE (Type II) have been visualized (Fig. 4). It is observed that when overall BWE values are small, WSE can better distinguish between normal samples and abnormal morphological samples. For instance, in patients 106, 200, 201, 210, 223, and 233, where the overall BWE values are widely spread (with many sample values concentrated between 2 and 4), using WSE alone cannot effectively differentiate abnormal morphology. Conversely, in patients 119, 208, 219, and 221, where overall BWE values range from 0 to 2, WSE can effectively distinguish abnormal morphological samples. This further underscores that when the amplitude patterns of normal and abnormal morphologies are similar, the introduction of phase information provides significant distinguishing information. However, when inherent amplitude pattern differences exist between normal and abnormal samples, the benefits of phase information from WSE for distinguishing positive samples will decrease. These visual results provide a basis for determining the applicability of amplitude and phase information and demonstrate the necessity of integrating both modes of information to form MEE. To further enhance the value of MEE, we attempt to provide a method for screening SVEBs, a rhythmic arrhythmia that lacks distinct morphological differences from SBs. Therefore, we suggest adapting the fluctuation rate threshold based on RR interval information, as previously attempted by researchers [6, 35]. We derive the threshold $\beta$ for screening supraventricular premature beats: $\beta = \frac{f(k)}{2 \times |RR_{k+1} - RR_k|/(RR_{k+1} + RR_k)}$, where $RR_k$ and $RR_{k+1}$ represent the pre-RR and post-RR intervals of the current heartbeat $k$, respectively. When SVEBs occur, the threshold $\beta$ will adaptively decrease to improve the detection rate of SVEBs. Through these discussions, we hope this research can bring new insights and methods to the field of unsupervised arrhythmia screening, providing lower-cost and more efficient auxiliary tools for clinical practice.

Moreover, we evaluate the feasibility of dataset pruning using MEE, selecting morphologically redundant negative samples to achieve class balance. Dataset pruning is crucial because negative samples significantly outnumber positive samples in real-world scenarios. In order to stabilize the learning of various classes by ML models, using the Synthetic Minority Over-sampling Technique (SMOTE) [36] algorithm would result in oversampling the number of positive samples to match the number of negative samples, thereby substantially increasing the training data volume. However, in economically underdeveloped regions, hardware resources remain a significant constraint on AI model training. Previous studies have shown that random pruning of negative samples to address class imbalance issues can yield considerable benefits in model training [37]. Nevertheless, random pruning still faces the risk of decreasing data representation richness. The experiments in our study (Fig. 9 and Table 3) demonstrate that MEE-guided pruning offers a more reasonable and economical approach to model training. Although MEE is calculated on a per-beat basis, our research suggests its potential application in both single-lead ECGs and standard 12-lead ECGs. In this paper, the computation for the 12-lead ECGs is conducted using a representative lead II. In the future, simultaneous analysis of multiple leads, such as Limb lead II and Chest lead V1, could enhance the computation of 12-lead ECGs by leveraging their ability to capture different aspects of cardiac activity. These leads provide complementary information from various angles and orientations, contributing to a more comprehensive understanding of the cardiac vector's projection in different planes. Through measurements using representative dual leads, we can further explore how to rapidly extend the MEE method to more comprehensive and global analyses of electrocardiographic signals, enhancing sensitivity to different cardiac angles and making it more practical.

The experiments also demonstrate the ability of MEE to identify noise regions (Fig. 10), indicating an interesting property of MEE: segments containing normal ECG morphology patterns consistently exhibit lower MEE values, whereas segments containing morphological arrhythmias or noise yield higher values. This proves the good adaptability of MEE to ECG morphology. Based on this, we further contemplate additional potential applications. Considering manual ECG annotation, professional annotators such as clinicians or data annotation companies may occasionally mislabel signals when faced with a large volume of data to annotate. We can utilize the MEE metric to select similar samples by analyzing metric values. If some values significantly differ among samples with the same label, we will consider that the data may not be correctly labeled but rather mislabeled. The mislabeled region can be identified so that professional annotators can re-examine the region label. This iterative annotation approach will alleviate misguidance to model training caused by noise labels.

Finally, we design an intuitive and user-friendly interface incorporating the proposed MEE metric, forming an auxiliary tool that clinicians can use in clinical practice (Fig. 12). Through detailed analysis of MEE values, we can identify abnormal heartbeats, diversity in morphologies of the same type of heartbeat, and noise. It provides displays of entropy measures at the beat level for BWE, WSE, and MEE, and user experience indicates that the interface is very smooth when analyzing beat-level indices, benefiting from the fast computation efficiency of MEE. Due to MEE's excellent descriptive capability for morphology, future considerations will also include fine-tuning to address various signals with physiological significance, such as photoplethysmography (PPG), ballistocardiography (BCG) signals, among others.

## 6. CONCLUSION

This paper attempts to develop an amplitude-phase fusion metric for ECG morphological analysis and introduces an entropy-based metric, MEE, for the first time. The experiments demonstrate the potential of this metric in the following applications: 1) Unsupervised anomaly pattern detection at the heartbeat level for long-term data from individual patients. 2) Optimization of morphological diversity for samples of the same type to obtain rich feature representations. 3) Sensitivity to noise. Based on these results, we develop a user-friendly interactive interface to facilitate rapid analysis of ECG data by relevant personnel. Additionally, the evaluation of noise resistance and sample processing time demonstrates its robustness and the potential for real-time monitoring. We believe that this intriguing metric, as proposed, will demonstrate greater practical value with further refinement and inspire researchers to fully integrate amplitude and phase patterns contained in ECG for the design of efficient and valuable metrics.